\documentstyle[pra,preprint,aps]{revtex}
\draft

\newcommand{\ba}{\begin{eqnarray}}
\newcommand{\ea}{\end{eqnarray}}
\begin{document}
\pagestyle{plain}
\def\ii{\'\i}

\title{A local anharmonic treatment of vibrations of methane}

\author{R. Lemus$^{1)}$, F. P\'erez-Bernal${^2)}$, A. Frank$^{1,3)}$, 
R. Bijker$^{1)}$ and J.M. Arias$^{2)}$}
\address{
$^{1)}$ Instituto de Ciencias Nucleares, 
        Universidad Nacional Aut\'onoma de M\'exico, 
        A.P. 70-543, 04510 M\'exico D.F., M\'exico\\
$^{2)}$ Departamento de F\'{\i}sica At\'omica, Molecular y Nuclear,
        Facultad de F\'{\i}sica, Universidad de Sevilla,
        Apdo. 1065, 41080 Sevilla, Espa\~na\\
$^{3)}$ Instituto de F\'{\i}sica, Laboratorio de Cuernavaca,
        A.P. 139-B, Cuernavaca, Morelos, M\'exico}
\maketitle

\begin{abstract}
The stretching and bending vibrations of methane are studied in a local 
anharmonic model of molecular vibrations. The use of symmetry-adapted 
operators reduces the eigenvalue problem to block diagonal form. 
For the 44 observed energies we obtain a fit with a standard 
deviation of 0.81 cm$^{-1}$ (and a r.m.s. deviation of 1.16
cm$^{-1}$).\\
\mbox{}
\mbox{}\\
Se hace una descripci\'on de las vibraciones de tensi\'on y 
flexi\'on de metano en el marco de un modelo local anarm\'onico de 
vibraciones mo\-le\-cu\-la\-res.  El uso de operadores y funciones adaptadas 
por simetr\ii a permite reducir el problema de valores propios a una 
forma diagonal.  Para las 44 energ\ii as observadas se obtiene un 
ajuste con una desviaci\'on est\'andar de 0.81 cm$^{-1}$ 
(correspondiente a una desviaci\'on cuadr\'atica media de 1.16 
cm$^{-1}$).\\
\mbox{}\\
Descriptores:  Vibraciones moleculares, modelo anarm\'onico, metano\\
\mbox{}
\end{abstract}

\section{Introduction}

The development and refinement of experimental techniques in 
high resolution spectroscopy has generated a wealth of new data 
on rovibrational spectra of polyatomic molecules. 
Highly symmetric molecules, such as tetrahedral XY$_4$ systems, 
form an ideal testing ground for models of molecular structure. 
On the one hand, the high degree of 
symmetry tends to reduce the complexity of the spectrum and on the 
other hand, the use of symmetry concepts and group theoretical 
techniques may help to interpret the data and eventually suggest 
new experiments \cite{Bobin}. A good example is the methane molecule, 
for which there exists a large amount of information on 
vibrational energies. 

Theoretically, the force field constants of methane from which the 
spectrum can be generated \cite{Hecht} can be obtained  
from anharmonic force field calculations \cite{GR} or 
{\it ab initio} calculations, see {\it e.g.} \cite{Raynes,Lee}. 
In a more empirical approach, the vibrational Hamiltonian is 
expressed in terms of curvilinear internal coordinates, which are 
symmetrized for the bending variables, but not for the stretching 
variables. The model parameters are optimized in a fit to experimental 
vibrational energies \cite{Halonen}. 

The aim of this paper is to investigate the stretching and bending 
vibrations of methane up to three quanta in a symmetry-adapted 
vibrational model. The method is based on the use of symmetrized 
(both for bending and stretching variables) internal 
coordinates. The fundamental idea is to incorporate anharmonic 
effects in the local modes from the outset. 
This is done by substituting the standard creation and annihilation 
operators associated with the internal coordinates by $SU(2)$ 
operators which carry the intrinsic anharmonicity of the local 
modes. The result can be viewed as a symmetrized (and in other ways 
improved) version of our previous work on methane \cite{LF}. 

\section{Local anharmonic symmetrized coordinates} 

The vibrational Hamiltonian of methane is usually expressed 
in terms of curvilinear internal coordinates $S_i$ and their 
conjugate momenta $P_i$ \cite{Hoy,WDC}.  
Both the kinetic and potential energy are then expanded as a 
Taylor series around the equilibrium configurations. 
In practice, it is convenient to introduce symmetry-adapted 
curvilinear internal coordinates and their conjugate momenta 
\cite{GR} 
\ba 
S_{x,\Gamma_{\gamma}} \;=\; \sum_{i=1}^{10} 
\alpha_{x,\Gamma_{\gamma}}^{i} \, S_{i} ~, \hspace{1cm}  
P_{i} \;=\; \sum_{x,\Gamma_{\gamma}} 
\alpha_{x,\Gamma_{\gamma}}^{i} \, P_{x,\Gamma_{\gamma}} ~, 
\label{cm} 
\ea
since it reduces the Hamiltonian to block diagonal form. 
In the case of methane there is a redundancy between the 
curvilinear coordinates which describe the bending degrees of 
freedom. The redundant or spurious combination can be removed 
from the outset by restricting the labels $\Gamma$ and $x$ to  
the fundamental modes only: $\Gamma=A_1$, $F_2$ for stretching 
vibrations $(x=s)$ and $\Gamma=E$, $F_2$ for bending vibrations $(x=b)$. 
However, as a consequence the inverse transformations 
of Eq.~(\ref{cm}) become nonlinear \cite{Hoy,WDC}. In the present 
discussion we prefer to keep the redundant coordinate (with 
symmetry labels $x=b$, $\Gamma=A_1$), and to remove the 
spurious states at a later stage. In this case the inverse 
relations are linear just as Eq.~(\ref{cm}) itself.  

Next we introduce instead of the symmetrized coordinates and momenta 
of Eq.~(\ref{cm}) creation and annihilation operators  
\ba 
\sqrt{\frac{\beta_{x,\Gamma}}{2\hbar}} \, S_{x,\Gamma_{\gamma}} 
- \frac{1}{\sqrt{2\beta_{x,\Gamma}\hbar}} 
\, i P_{x,\Gamma_{\gamma}} &=& b^{\dagger}_{x,\Gamma_{\gamma}} ~,
\nonumber\\
\sqrt{\frac{\beta_{x,\Gamma}}{2\hbar}} \, S_{x,\Gamma_{\gamma}} 
+ \frac{1}{\sqrt{2\beta_{x,\Gamma}\hbar}} 
\, i P_{x,\Gamma_{\gamma}} &=& b_{x,\Gamma_{\gamma}} ~. 
\ea
The coefficients $\beta_{x,\Gamma}$ represent scaling factors 
for each vibrational mode. According to the discussion following 
Eq.~(\ref{cm}) the symmetrized operators can be expanded in terms 
of local operators as 
\ba 
b^{\dagger}_{x,\Gamma_{\gamma}} \;=\; \sum_{i=1}^{10} 
\alpha_{x,\Gamma_{\gamma}}^{i} \, b^{\dagger}_{i} ~, \hspace{1cm} 
b_{x,\Gamma_{\gamma}} \;=\; \sum_{i=1}^{10} 
\alpha_{x,\Gamma_{\gamma}}^{i} \, b_{i} ~. \label{bsymm}
\ea 

As mentioned before, the basic idea of the present approach is to 
explicitly incorporate anharmonic effects in the local modes. It has 
been shown that the anharmonicities induced by potentials such as the 
Morse and P\"oschl-Teller oscillators can be described in terms of 
$U(2)$ algebras \cite{AGI,vibron}. Hereto we construct an anharmonic 
representation of the local operators $b^{\dagger}_i$ and $b_i$ through 
the correspondence \cite{Be4,X3} 
\ba
b^{\dagger}_i \;\rightarrow\; a^{\dagger}_i 
\;\equiv\; \frac{\hat{J}_{-,i}}{\sqrt{N_i}} ~, \hspace{1cm} 
b_i \;\rightarrow\; a_i 
\;\equiv\; \frac{\hat{J}_{+,i}}{\sqrt{N_i}} ~, 
\label{anharm}
\ea
where $\hat J_i^2 = \hat J_{0,i}^2 + (\hat J_{+,i} \hat J_{-,i} 
+ \hat J_{-,i} \hat J_{+,i})/2 = \hat N_i(\hat N_i+2)/4$, {\it i.e.} 
$N_i/2=j_i$ \cite{X3}. 
The anharmonic operators satisfy the commutation relation
\ba
\, [ a_i, a_j^{\dagger} ] &=&  
\frac{1}{\sqrt{N_iN_j}} [ \hat J_{+,i}, \hat J_{-,j} ]
\;=\; \delta_{ij} \, \frac{2 \hat J_{0,i}}{N_i} 
\;=\; \delta_{ij} \left( 1 - \frac{2 \hat v_i}{N_i} \right) ~,
\nonumber\\
\, [ a_i, \hat v_j ] &=& \frac{1}{\sqrt{N_i}} 
[ \hat J_{+,i}, \frac{\hat N_j}{2} - \hat J_{0,j} ]
\;=\; \delta_{ij} \, \frac{\hat J_{+,i}}{\sqrt{N_i}} 
\;=\; \delta_{ij} \, a_i ~, 
\nonumber\\
\, [ a_i^{\dagger}, \hat v_j ] &=& \frac{1}{\sqrt{N_i}} 
[ \hat J_{-,i}, \frac{\hat N_j}{2} - \hat J_{0,j} ]
\;=\;-\delta_{ij} \, \frac{\hat J_{-,i}}{\sqrt{N_i}} 
\;=\;-\delta_{ij} \, a_i^{\dagger} ~. 
\label{comm} 
\ea
The operators $J_{\mu,i}$ with $\mu=\pm,0$ together with 
the number operator $\hat N_i$ are the generators of $U_i(2)$. 
In the limit $N_i \rightarrow \infty$ we recover the harmonic 
description in terms of $b_i^{\dagger}$, $b_i$ and 
$\hat v_i = b_i^{\dagger} b_i$. 
With each internal coordinate we associate a 
$U(2)$ algebra. For the methane molecule this leads to 
four $U(2)$ algebras for the stretching degrees of freedom 
and six more for the bending degrees of freedom (of the latter, 
one linear combination is spurious). 
The molecular dynamical group is then given by 
\ba
{\cal G} &=& U_1(2) \otimes U_2(2) \otimes \ldots \otimes U_{10}(2) ~.
\label{group}
\ea
The Hamiltonian and other operators of interest are expressed in terms 
of the generators $\hat J_{\mu,i}$, $\hat N_i$ of the $U_i(2)$ algebras 
in Eq.~(\ref{group}). The local basis states for each oscillator are 
$|N_i,v_i \rangle$, where $v_i=0,1, \ldots [N_i/2]$ 
denotes the number of oscillator quanta in the $i$-th oscillator 
and $N_i$ is related to the depth of the anharmonic potential 
\cite{vibron,thebook,X3}. 

For the CH$_4$ molecule there are two different boson numbers: 
$N_s$ for the stretching modes and $N_b$ for the bending modes.
The tetrahedral symmetry of methane is taken into account by 
symmetrizing the local operators $\hat J_{\mu,i}$ \cite{LF}
\ba
\hat T_{\mu,x}^{\Gamma_{\gamma}} \;=\; \sum_{i=1}^{10} 
\alpha_{x,\Gamma_{\gamma}}^{i} \, \hat J_{\mu,i} ~. 
\label{tdt}
\ea 
The coefficients $\alpha^i_{\Gamma_{\gamma},x}$ are the same as in 
Eq.~(\ref{cm}). The symmetrized tensor operators of Eq.~(\ref{tdt}) 
correspond to ten degrees of freedom, four of which 
($A_1 \oplus F_2$) are related to stretching modes and six 
($A_1 \oplus E \oplus F_2$) to the bendings. Consequently we can 
identify the tensor with $x=b$ and $\Gamma=A_1$ as the operator associated 
to a spurious mode. This identification makes it possible to eliminate the 
spurious states {\em exactly}. This is achieved by (i) ignoring the 
$\hat{T}_{\mu,b}^{A_1}$ tensor in the construction of the Hamiltonian, 
and (ii) diagonalizing this Hamiltonian in a symmetry-adapted basis from 
which the spurious mode has been removed \cite{comment,preprint}.
It is important to note that, although in general in the presence of 
anharmonic interactions only approximate methods can be developed to 
eliminate spurious degrees of freedom, the particular anharmonization 
provided by $U(2)$ admits a symmetry procedure to exclude 
the unphysical states exactly \cite{preprint}.

\section{The vibrational Hamiltonian}

The vibrational Hamiltonian for methane can now be constructed by 
repeated couplings of the tensors of Eq.~(\ref{tdt}) to a scalar 
($\Gamma=A_1$) under the tetrahedral group ${\cal T}_d$. We use the 
standard labelling for the vibrational basis states: 
$(\nu_1 \nu_2^{l_2} \nu_3^{l_3} \nu_4^{l_4})$, where 
$\nu_1$, $\nu_3$ and $\nu_2$, 
$\nu_4$ denote the number of quanta in the $A_1$, $F_2$ stretching 
modes, and in the $E$, $F_2$ bending modes, respectively. 
The labels $l_i$ are related to the vibrational angular momentum 
associated with degenerate vibrations. The allowed values are 
$l_i=\nu_i,\nu_i-2,\ldots,1$ or 0 for $\nu_i$ odd or even \cite{Herzberg2}.

In this paper, the Hamiltonian is taken to be diagonal in the total 
number of quanta $V=\nu_1+\nu_2+\nu_3+\nu_4$, and in the polyad
$V^\prime =2\nu_1 + \nu_2 + 2\nu_3 + \nu_4$, but 
 does not contain 
explicit Fermi interactions between the stretching and bending vibrations. 
Fermi interactions can be included in the model by means of
additional interactions 
which are diagonal in the polyad $V'=2\nu_1+\nu_2+2\nu_3+\nu_4$ and
exchange quanta between the stretching and bending modes.  In that
case the number of quanta $V$ ceases to be conserved.  

According to the above procedure, we now construct the ${\cal T}_d$ 
invariant interactions that are at most quadratic in the generators 
and conserve the total number of quanta 
\ba
\hat{\cal H}_{x,\Gamma} &=& \frac{1}{2N_{x}} \sum_{\gamma} \left( 
  \hat T^{\Gamma_{\gamma}}_{-,x} \, \hat T^{\Gamma_{\gamma}}_{+,x}
+ \hat T^{\Gamma_{\gamma}}_{+,x} \, \hat T^{\Gamma_{\gamma}}_{-,x} 
\right) ~,
\nonumber\\
\hat{\cal V}_{x,\Gamma} &=& \frac{1}{N_{x}} \sum_{\gamma} \,
\hat T^{\Gamma_{\gamma}}_{0,x} \, \hat T^{\Gamma_{\gamma}}_{0,x} ~.
\label{hv}
\ea
Here $\Gamma=A_1$, $F_2$ for the stretching vibrations $x=s$ and 
$\Gamma=E$, $F_2$ for the bending vibrations $x=b$. In addition to
Eq.~(\ref{hv}), there are two stretching-bending interactions
\ba
\hat{\cal H}_{sb} &=& \frac{1}{2\sqrt{N_sN_b}} \sum_{\gamma} \left( 
  \hat T^{F_{2\gamma}}_{-,s} \, \hat T^{F_{2\gamma}}_{+,b}
+ \hat T^{F_{2\gamma}}_{+,s} \, \hat T^{F_{2\gamma}}_{-,b} \right) ~,
\nonumber\\
\hat{\cal V}_{sb} &=& \frac{1}{\sqrt{N_sN_b}} \sum_{\gamma} \, 
\hat T^{F_{2\gamma}}_{0,s} \, \hat T^{F_{2\gamma}}_{0,b} ~.
\label{hvsb}
\ea
The zeroth order vibrational Hamiltonian is now written as
\ba
\hat H_0 &=& \omega_1 \, \hat{\cal H}_{s,A_1} 
   + \omega_2 \, \hat{\cal H}_{b,E} 
   + \omega_3 \, \hat{\cal H}_{s,F_2} 
   + \omega_4 \, \hat{\cal H}_{b,F_2} 
   + \omega_{34} \, \hat{\cal H}_{sb} 
\nonumber\\
&& + \alpha_2 \, \hat{\cal V}_{b,E}  
   + \alpha_3 \, \hat{\cal V}_{s,F_2} 
   + \alpha_4 \, \hat{\cal V}_{b,F_2} 
   + \alpha_{34} \, \hat{\cal V}_{sb} ~. \label{h0}
\ea
The interaction terms of $\hat H_0$ can be rewritten in terms of the 
Casimir operators of subgroups of Eq.~(\ref{group}) which were used 
in \cite{LF}. The interaction $\hat{\cal V}_{A_{1,s}}$ has not been 
included since the combination
\ba
\sum_{\Gamma=A_1,F_2} \left( \hat{\cal H}_{s,\Gamma} 
+ \hat{\cal V}_{s,\Gamma} \right) &=& \frac{1}{4N_s}
\sum_{i=1}^{4} \hat N_i(\hat N_i+2) ~,
\ea
corresponds to a constant $N_s+2$. A similar relation holds for the 
bending interactions, but in this case the interaction 
$\hat{\cal V}_{b,A_1}$ has already been excluded in order to remove 
the spurious $A_1$ bending mode. The subscripts of the parameters 
correspond to the $(\nu_1 \nu_2^{l_2} \nu_3^{l_3} \nu_4^{l_4})$ 
labeling of a set of basis states for the vibrational levels of CH$_4$. 

In the harmonic limit the interactions of Eqs.~(\ref{hv}) 
and~(\ref{hvsb}) attain a particularly simple form, which can be 
directly related to configuration space interactions 
\cite{Hecht,Halonen}. This limit is obtained by interpreting 
Eq.~(\ref{anharm}) in the opposite sense \cite{Be4,X3}, and 
corresponds group theoretically to the contraction of the $SU(2)$ 
algebra to the Weyl algebra. In the harmonic limit the interactions 
of Eqs.~(\ref{hv}) and~(\ref{hvsb}) can be expressed in terms of the 
symmetrized harmonic operators of Eq.~(\ref{bsymm}) 
\ba
\lim_{N_{x} \rightarrow \infty} \, \hat{\cal H}_{x,\Gamma} 
&=& \frac{1}{2} \sum_{\gamma} \left( 
  b^{\dagger}_{x,\Gamma_{\gamma}} \, b_{x,\Gamma_{\gamma}} 
+ b_{x,\Gamma_{\gamma}} \, b^{\dagger}_{x,\Gamma_{\gamma}} \right) ~,
\nonumber\\
\lim_{N_{x} \rightarrow \infty} \, \hat{\cal V}_{x,\Gamma} &=& 0 ~,
\nonumber\\
\lim_{N_s, N_b \rightarrow \infty} \, \hat{\cal H}_{sb} 
&=& \frac{1}{2} \sum_{\gamma} \left( 
  b^{\dagger}_{s,F_{2\gamma}} \, b_{b,F_{2\gamma}} 
+ b_{s,F_{2\gamma}} \, b^{\dagger}_{b,F_{2\gamma}} \right) ~,
\nonumber\\
\lim_{N_s, N_b \rightarrow \infty} \, \hat{\cal V}_{sb} &=& 0 ~.
\label{harlim}
\ea
From Eq.~(\ref{harlim}) we find a direct physical interpretation for 
the interaction terms. The $\hat{\cal H}_{x,\Gamma}$ terms represent 
the anharmonic counterpart of the harmonic interactions, while 
the $\hat{\cal V}_{x,\Gamma}$ terms are purely anharmonic contributions 
which vanish in the harmonic limit. 

The zeroth order Hamiltonian of Eq.~(\ref{h0}), however, is not sufficient 
to obtain a high-quality fit of the vibrations of methane (see also 
\cite{Halonen}. The use of symmetrized operators of Eq.~(\ref{tdt}) 
makes it possible to construct higher order (quartic) terms in a 
straightforward and systematic way. For the study of the vibrational 
excitations of methane we propose to use the following 
${\cal T}_d$ invariant quartic Hamiltonian 
\ba
\hat H &=& \omega_1 \, \hat{\cal H}_{s,A_1}
         + \omega_2 \, \hat{\cal H}_{b,E} 
         + \omega_3 \, \hat{\cal H}_{s,F_2} 
         + \omega_4 \, \hat{\cal H}_{b,F_2} 
         + \alpha_3 \, \hat{\cal V}_{s,F_2}
\nonumber\\
&& + X_{11} \left( \hat{\cal H}_{s,A_1} \right)^2
   + X_{22} \left( \hat{\cal H}_{b,E  } \right)^2
   + X_{33} \left( \hat{\cal H}_{s,F_2} \right)^2
   + X_{44} \left( \hat{\cal H}_{b,F_2} \right)^2
\nonumber\\
&& + X_{12} \left( \hat{\cal H}_{s,A_1} \, \hat{\cal H}_{b,E  } \right)
   + X_{14} \left( \hat{\cal H}_{s,A_1} \, \hat{\cal H}_{b,F_2} \right)
\nonumber\\
&& + X_{23} \left( \hat{\cal H}_{b,E  } \, \hat{\cal H}_{s,F_2} \right) 
   + X_{24} \left( \hat{\cal H}_{b,E  } \, \hat{\cal H}_{b,F_2} \right) 
   + X_{34} \left( \hat{\cal H}_{s,F_2} \, \hat{\cal H}_{b,F_2} \right) 
\nonumber\\
&& + g_{22} \, \left( \hat l_{A_2} \right)^2  
   + g_{33} \, \sum_{\gamma} \hat l^{F_1}_{s,\gamma} \, 
                             \hat l^{F_1}_{s,\gamma}
   + g_{44} \, \sum_{\gamma} \hat l^{F_1}_{b,\gamma} \, 
                             \hat l^{F_1}_{b,\gamma} 
   + g_{34} \, \sum_{\gamma} \hat l^{F_1}_{s,\gamma} \, 
                             \hat l^{F_1}_{b,\gamma} 
\nonumber\\
&& + t_{33} \, \hat{\cal O}_{ss}
   + t_{44} \, \hat{\cal O}_{bb}
   + t_{34} \, \hat{\cal O}_{sb}
   + t_{23} \, \hat{\cal O}_{2s}
   + t_{24} \, \hat{\cal O}_{2b} ~. \label{hamilt}
\ea
Each one of the interaction terms of the Hamiltonian of Eq.~(\ref{hamilt}) 
has a direct physical interpretation and a specific action on the 
various modes. The $\omega_i$ and $\alpha_3$ terms have already been 
discussed in Eq.~(\ref{harlim}). The $X_{ij}$ terms are quadratic in the 
operators $\hat{\cal H}_{x,\Gamma}$ and hence represent anharmonic 
vibrational interactions. The $g_{ij}$ terms are related to the 
vibrational angular momenta associated with the degenerate vibrations 
\cite{Hecht} and give rise to a splitting of vibrational levels with 
the same values of $(\nu_1 \nu_2 \nu_3 \nu_4)$ but with different 
$l_2$, $l_3$ and/or $l_4$. They can be expressed in terms of the 
symmetrized tensors of Eq.~(\ref{tdt}) as 
\ba 
\hat l^{A_2} &=& -i \, \sqrt{2} \frac{1}{N_b} 
[ \hat T^{E}_{-,b} \times \hat T^{E}_{+,b} ]^{A_2} ~,
\nonumber\\
\hat l^{F_{1\gamma}}_x &=& +i \, \sqrt{2} \frac{1}{N_x} 
[ \hat T^{F_2}_{-,x} \times \hat T^{F_2}_{+,x} ]^{F_{1\gamma}} ~.
\label{gij}
\ea
The square brackets in Eq.~(\ref{gij}) denote the tensor coupling 
under the point group ${\cal T}_d$ \cite{LF}.
In the harmonic limit, the expectation value of the $\omega_i$, 
$\alpha_3$, $X_{ij}$ and $g_{ij}$ terms in Eq.~(\ref{hamilt}) leads 
to the familiar Dunham expansion \cite{Herzberg2} 
\ba
\sum_i \omega_i \, (v_i + \frac{d_i}{2}) + \sum_{i \leq j} 
X_{ij} \, (v_i + \frac{d_i}{2}) (v_j + \frac{d_j}{2})
+ \sum_{i \leq j} g_{ij} \, l_i l_j ~, \label{Dunham}
\ea
where $d_i$ is the degeneracy of the vibration. The $t_{ij}$ terms are 
quartic operators of the type discussed by Hecht \cite{Hecht} 
and give rise to further splittings of the vibrational levels 
$(\nu_1 \nu_2 \nu_3 \nu_4)$ into its possible sublevels. 
They can be expressed in terms of the tensor operators of 
Eq.~(\ref{tdt}) as
\ba
\hat{\cal O}_{xy} &=& \frac{1}{N_x N_y}
\left( 6 \sum_{\gamma} 
[ \hat T^{F_2}_{-,x} \times \hat T^{F_2}_{-,y} ]^{E_{\gamma}} \,
[ \hat T^{F_2}_{+,y} \times \hat T^{F_2}_{+,x} ]^{E_{\gamma}} \right. 
\nonumber\\
&& \hspace{3cm} \left. -4 \sum_{\gamma}
[ \hat T^{F_2}_{-,x} \times \hat T^{F_2}_{-,y} ]^{F_{2\gamma}} \,
[ \hat T^{F_2}_{+,y} \times \hat T^{F_2}_{+,x} ]^{F_{2\gamma}} \right) ~, 
\nonumber\\
\hat{\cal O}_{2x} &=& \frac{1}{N_b N_x} 
\left( 8 \sum_{\gamma} 
[ \hat T^E_{-,b} \times \hat T^{F_2}_{-,x} ]^{F_{1\gamma}} \,
[ \hat T^E_{+,b} \times \hat T^{F_2}_{+,x} ]^{F_{1\gamma}} \right.
\nonumber\\
&& \hspace{3cm} \left. -8 \sum_{\gamma}
[ \hat T^E_{-,b} \times \hat T^{F_2}_{-,x} ]^{F_{2\gamma}} \,
[ \hat T^E_{+,b} \times \hat T^{F_2}_{+,x} ]^{F_{2\gamma}} \right) ~.
\label{tij}
\ea
In the harmonic limit the $t_{ij}$ terms have the same interpretation 
as in \cite{Hecht,Halonen}. The $\hat{\cal O}_{ss}$, $\hat{\cal O}_{bb}$ 
and $\hat{\cal O}_{sb}$ terms give rise to a splitting of the $E$ and 
$F_2$ vibrations belonging to the 
$(\nu_1 \nu_2^{l_2} \nu_3^{l_3} \nu_4^{l_4})=(0 0^0 2^2 0^0)$, 
$(0 0^0 0^0 2^2)$ and $(0 0^0 1^1 1^1)$ levels, respectively. 
Similarly, the $\hat{\cal O}_{2s}$ and $\hat{\cal O}_{2b}$ 
terms split the $F_1$ and $F_2$ vibrations belonging to the 
$(0 1^1 1^1 0^0)$ and $(0 1^1 0^0 1^1)$ overtones, respectively. 

We remark that, whereas the $\omega_i$, $\alpha_3$ and $X_{ij}$ 
terms can be rewritten in terms of the Casimir invariants of \cite{LF} 
and products thereof, the $g_{ij}$ and $t_{ij}$ terms cannot be expressed 
in this way. These interactions involve intermediate couplings 
with $\Gamma=A_2$, $F_1$, $E$, $F_2$ symmetry, that are not symmetric 
($\Gamma=A_1$) as is the case for the invariant operators.   

\section{Results}

The Hamiltonian of Eq.~(\ref{hamilt}) involves 23 interaction 
strengths and the two boson numbers, $N_s$ and $N_b$. The vibron 
number associated with the stretching vibrations is determined 
from the spectroscopic constants $\omega_e$ and $x_e \omega_e$ 
for the CH molecule to be $N_s=43$ \cite{vibron,LF}. The vibron number 
for the bending vibrations, which are far more harmonic than the 
stretching vibrations, is taken to be $N_b=150$. We have carried out 
a least-square fit to the vibrational spectrum of methane including 
44 experimental energies from \cite{Halonen}, 
\cite{Champion}--\cite{Margolis} with equal weights. 

The values of the fitted parameters are presented in the second column 
of Table~\ref{fit} (Fit 1). In Table~\ref{CH4} we compare the results 
of our calculation with the experimentally observed energies. All 
predicted levels up to $V=\nu_1+\nu_2+\nu_3+\nu_4=3$ quanta are included. 
The quality of the fit is expressed in terms of the r.m.s. deviation 
\ba
\delta &=& \left[ \sum_{i=1}^n 
(E_{exp}^i-E_{cal}^i)^2/(N_{exp}-N_{par}) \right]^{1/2} ~,
\label{rms}
\ea
where $N_{exp}$ is the total number of experimental energies 
and $N_{par}$ the number of parameters used in the fit. 
We find a good overall fit to the observed levels with a r.m.s. 
deviation of $\delta = 1.16$ cm$^{-1}$ for 44 energies 
(and a standard deviation of $\sigma = 0.81$ cm$^{-1}$). 
The deviations with experiment are fairly constant over the 
entire energy range up to 9000 cm$^{-1}$, the largest one being 
$\Delta E=-2.22$ cm$^{-1}$. 
A statistical analysis of the error in the parameters ({\em i.e.} 
the variation in a given parameter such that the r.m.s. does not 
increase more than a certain fraction) shows that the fitted 
parameters are well determined. 
The cross-anharmonicity $X_{13}$ was not included in the fit, due to 
lack of data for the $(\nu_1 0 \nu_3 0)$ vibrations. 

A particularly important role is played by the $\alpha_3$ term. 
Eq.~(\ref{comm}) shows that this type of terms can be rewritten as 
particular higher order interactions in the operators $\hat J_{\pm,i}$. 
In order to address the importance of this term, we have 
carried out a fit in the harmonic limit ($N_s \rightarrow \infty$, 
$N_b \rightarrow \infty$). In this limit the $\alpha_3$ term 
vanishes and the Hamiltonian of Eq.~(\ref{hamilt}) is equivalent 
to the vibrational Hamiltonian of \cite{Hecht}, 
the harmonic frequencies $\omega_i$ and anharmonic constants 
$X_{ij}$, $g_{ij}$ and $t_{ij}$ having the same meaning. 
A comparison between the parameter values and the r.m.s. deviations 
of Fits 1 and 2 in Table~\ref{fit} shows that the 
$\alpha_3$ term and the anharmonic effects in the interaction 
terms of Eq.~(\ref{hamilt}) can only be compensated for in part by the 
anharmonicity constants $X_{ij}$. The r.m.s. deviation increases 
from $\delta = 1.16$ cm$^{-1}$ for Fit 1 to 
$\delta = 20.42$ cm$^{-1}$ for Fit 2. 

For comparison we show in Table~\ref{calcs} the results of some 
other recent model calculations. 

\section{Summary and conclusions}

In summary, in this paper we have studied the vibrational excitations 
of methane in a model based on the use of symmetry-adapted 
internal coordinates, in which anharmonic effects 
are introduced in the local modes. We find an overall fit 
to the 44 observed levels with a r.m.s. deviation of 
$\delta = 1.16$ cm$^{-1}$ 
(and a standard deviation of $\sigma = 0.81$ cm$^{-1}$),  
which can be considered of spectroscopic quality. 
We pointed out that the $\alpha_3$ term in combination with 
the anharmonic effects in the other interaction terms plays an 
important role in obtaining a fit of this quality. 
Physically, these contributions arise from the anharmonic character 
of the local modes. They play an important role to describe 
the anharmonicities, especially for higher number of quanta. 
This conclusion is supported by earlier applications of this 
model to the Be$_4$ cluster \cite{Be4,Oax}, 
the H$_3^+$, Be$_3$ and Na$_3^+$ molecules \cite{X3}, and two 
isotopes of the ozone molecule \cite{ozone}. 

The present calculations only include interaction terms that 
are simultaneously diagonal in the total number of quanta
$V=\nu_1+\nu_2+\nu_3+\nu_4$,  and in the polyad $V^\prime = 2 (\nu_1
+ \nu_3) + \nu_2 + \nu_4$, but do 
 not contain explicit Fermi interactions between the 
stretching and bending vibrations. It is interesting to note that 
despite the absence of these interactions which are generally considered 
to be necessary for an adequate description of methane, we do obtain a 
high quality fit. Fermi interactions can be included in the present 
model by constructing a Hamiltonian which is still diagonal in the polyad 
$V'=2\nu_1+\nu_2+2\nu_3+\nu_4$ but that mixes the stretching and
bending modes \cite{vnueve}. It is well-known that energies 
only are not sufficient to distinguish between various model 
calculations. Other quantities, such as infrared and Raman transitions 
or Franck-Condon factors are more sensitive to details in the 
wave functions than energies, and provide a better test of 
different models of molecular structure. 
Work along these directions is in progress \cite{vnueve}. 

\section*{Acknowledgements}

This work was supported in part by DGAPA-UNAM under project IN101997, 
and the European Community under contract nr. CI1$^{\ast}$-CT94-0072.

\clearpage
\begin{table}
\centering
\caption[]{\small
Parameters in cm$^{-1}$ obtained in the fit to the vibrational 
energies of CH$_4$. The last column shows the results in the 
harmonic limit ($N_s \rightarrow \infty$, $N_b \rightarrow \infty$). 
\normalsize}
\vspace{15pt} \label{fit}
\begin{tabular}{c|rr}
\hline
& & \\
Parameter~~~ & Fit 1 & Fit 2 \\ 
& & \\
\hline
& & \\
$N_s$      &      43 & $\infty$ \\
$N_b$      &     150 & $\infty$ \\
$\omega_1$ & 2977.60 & 2967.40  \\
$\omega_2$ & 1554.83 & 1558.38  \\ 
$\omega_3$ & 3076.45 & 3081.34  \\
$\omega_4$ & 1332.22 & 1337.51  \\
$\alpha_3$ &  582.87 &      --  \\
$X_{11}$   &    3.69 & --21.30  \\ 
$X_{22}$   &    1.30 &  --1.17  \\ 
$X_{33}$   &    5.43 & --10.79  \\ 
$X_{44}$   &  --3.47 &  --6.26  \\ 
$X_{12}$   &  --3.60 &  --3.39  \\
$X_{13}$   &      -- &      --  \\
$X_{14}$   &  --2.86 &  --3.10  \\
$X_{23}$   & --11.14 &  --7.97  \\
$X_{24}$   &    1.00 &  --5.37  \\
$X_{34}$   &  --5.60 &  --3.46  \\
$g_{22}$   &  --0.46 &    0.37  \\
$g_{33}$   &    0.19 &  --4.35  \\
$g_{44}$   &    4.07 &    4.98  \\
$g_{34}$   &  --0.65 &  --0.74  \\
$t_{33}$   &    0.40 &  --1.25  \\
$t_{44}$   &    1.00 &    0.56  \\
$t_{34}$   &    0.21 &    0.24  \\
$t_{23}$   &  --0.39 &  --0.39  \\
$t_{24}$   &    0.13 &    0.91  \\
& & \\
\hline
& & \\
r.m.s. & 1.16 & 20.42 \\
& & \\
\hline
\end{tabular}

\end{table}

\clearpage
\begin{table}
\centering
\caption[]{\small
Fit to vibrational excitations of CH$_4$. The values of the parameters 
are given in the second column of Table\ref{fit}. Here 
$\Delta E=E_{cal}-E_{exp}$. The experimental energies are taken from 
\cite{Halonen}, \cite{Champion}--\cite{Margolis}.
The levels marked with an asterisk are taken from \cite{Herzberg1}, but 
were not included in the fit. 
The wave numbers are given in cm$^{-1}$. 
\normalsize}
\footnotesize
\vspace{15pt} \label{CH4}
\begin{tabular}{lcllr|lcllr}
\hline
& & & & & & & & & \\
$\Gamma$ & $(\nu_1 \nu_2 \nu_3 \nu_4)$ & 
$E_{cal}$ & $E_{exp}$ & $\Delta E$~~ & 
$\Gamma$ & $(\nu_1 \nu_2 \nu_3 \nu_4)$ & 
$E_{cal}$ & $E_{exp}$ & $\Delta E$ \\
& & & & & & & & & \\
\hline
& & & & & & & & & \\
$A_1$ & (1000) & 2916.32 & 2916.48 & --0.16~~ & 
      & (0111) & 5844.98 &         &        \\
$E$   & (0100) & 1533.46 & 1533.33 &   0.13~~ &
      & (1200) & 5974.81 &         &        \\
$F_2$ & (0001) & 1309.86 & 1310.76 & --0.90~~ &
      & (1011) & 7147.49 &         &        \\
      & (0010) & 3018.09 & 3019.49 & --1.40~~ &
      & (0021) & 7303.38 &         &        \\
      &        &         &         &        &
      & (2100) & 7315.60 &         &        \\
$A_1$ & (0002) & 2587.77 & 2587.04 &   0.73~~ &
      & (0120) & 7479.48 &         &        \\
      & (0200) & 3063.66 & 3063.65 &   0.01~~ &
      & (0120) & 7557.17 &         &        \\
      & (0011) & 4323.81 & 4322.72 &   1.09~~ &
      & (1020) & 8833.05 &         &        \\
      & (2000) & 5790.13 & 5790    &   0.13~~ &
$F_1$ & (0003) & 3920.46 & 3920.50 & --0.04 \\
      & (0020) & 5966.57 & 5968.1  & --1.53~~ &
      & (0102) & 4128.38 & 4128.57 & --0.19 \\
$E$   & (0002) & 2624.14 & 2624.62 & --0.48~~ &
      & (0201) & 4364.39 & 4363.31 &   1.08 \\
      & (0200) & 3065.22 & 3065.14 &   0.08~~ &
      & (0012) & 5620.08 &         &        \\
      & (0011) & 4323.09 & 4322.15 &   0.94~~ &
      & (0012) & 5630.76 &         &        \\
      & (1100) & 4446.41 & 4446.41 &   0.00~~ &
      & (1101) & 5755.58 &         &        \\
      & (0020) & 6045.03 & 6043.8  &   1.23~~ &
      & (0111) & 5829.79 &         &        \\
$F_1$ & (0101) & 2845.35 & 2846.08 & --0.73~~ &
      & (0111) & 5848.94 &         &        \\
      & (0011) & 4323.15 & 4322.58 &   0.57~~ &
      & (0210) & 6061.57 &         &        \\
      & (0110) & 4537.57 & 4537.57 &   0.00~~ &
      & (1011) & 7147.53 &         &        \\
$F_2$ & (0002) & 2612.93 & 2614.26 & --1.33~~ &
      & (0021) & 7303.29 &         &        \\
      & (0101) & 2830.61 & 2830.32 &   0.29~~ &
      & (0021) & 7343.21 &         &        \\
      & (1001) & 4223.46 & 4223.46 &   0.00~~ &
      & (1110) & 7361.79 &         &        \\
      & (0011) & 4321.02 & 4319.21 &   1.81~~ &
      & (0120) & 7518.70 &         &        \\
      & (0110) & 4543.76 & 4543.76 &   0.00~~ &
      & (0030) & 8947.65 & 8947.95 & --0.30 \\
      & (1010) & 5845.53 &         &        &
$F_2$ & (0003) & 3871.29 & 3870.49 &   0.80 \\
      & (0020) & 6003.65 & 6004.65 & --1.00~~ &
      & (0003) & 3931.36 & 3930.92 &   0.44 \\
      &        &         &         &        &
      & (0102) & 4143.09 & 4142.86 &   0.23 \\
$A_1$ & (0003) & 3909.20 & 3909.18 &   0.02~~ &
      & (0201) & 4349.01 & 4348.77 &   0.24 \\
      & (0102) & 4131.92 & 4132.99 & --1.07~~ &
      & (0201) & 4378.38 & 4379.10 & --0.72 \\
      & (0300) & 4595.26 & 4595.55 & --0.29~~ &
      & (1002) & 5523.80 &         &        \\
      & (1002) & 5498.66 &         &        &
      & (0012) & 5594.92 & 5597.14 & --2.22 \\
      & (0012) & 5617.16 &         &        &
      & (0012) & 5620.68 &         &        \\
      & (0111) & 5836.11 &         &        &
      & (0012) & 5632.36 &         &        \\
      & (1200) & 5973.26 &         &        &
      & (1101) & 5740.86 &         &        \\
      & (1011) & 7147.56 &         &        &
      & (0111) & 5830.28 &         &        \\
      & (0021) & 7300.85 &         &        &
      & (0111) & 5848.46 &         &        \\
      & (0120) & 7562.91 &         &        &
      & (0210) & 6054.58 &         &        \\
      & (3000) & 8583.81 &         &        &
      & (0210) & 6067.03 &         &        \\
      & (1020) & 8727.97 &         &        &
      & (2001) & 7094.16 &         &        \\
      & (0030) & 8975.64 & 8975.34 &   0.30~~ &
      & (1011) & 7145.84 &         &        \\
$A_2$ & (0102) & 4161.52 & 4161.87 & --0.35~~ &
      & (0021) & 7266.11 &         &        \\
      & (0300) & 4595.28 & 4595.32 & --0.04~~ &
      & (0021) & 7303.38 &         &        \\
      & (0111) & 5844.61 &         &        &
      & (0021) & 7344.87 &         &        \\
      & (0120) & 7550.53 &         &        &
      & (1110) & 7365.83 &         &        \\
$E$   & (0102) & 4105.22 & 4105.15 &   0.07~~ &
      & (0120) & 7514.67 & 7514$^{\ast)}$    &       \\
      & (0102) & 4152.15 & 4151.22 &   0.93~~ &
      & (2010) & 8594.90 & 8604$^{\ast)}$    &       \\
      & (0300) & 4592.13 & 4592.03 &   0.10~~ &
      & (1020) & 8786.05 & 8807$^{\ast)}$    &       \\
      & (1002) & 5535.04 &         &        &
      & (0030) & 8907.91 & 8906.78 &   1.13 \\
      & (0012) & 5620.36 &         &        &
      & (0030) & 9045.36 & 9045.92 & --0.56 \\
      & (0111) & 5836.45 &         &        &
      &        &         &         &        \\
& & & & & & & & & \\
\hline
\end{tabular}
\normalsize
\end{table}

\clearpage
\begin{table}
\centering
\caption[]{\small 
Standard and r.m.s. deviations in cm$^{-1}$ of some recent 
calculations of vibrational energies of CH$_4$.
\normalsize}
\vspace{15pt} \label{calcs}
\begin{tabular}{l|cccr}
\hline
& & & & \\
Reference & $N_{exp}$ & $N_{par}$ & $\sigma$ & $\delta$~~ \\
& & & & \\
\hline
& & & & \\
Lemus and Frank \cite{LF}~~ & 19 &  8 & 9.50 & 12.16 \\
Ma et al \cite{Ma}          & 19 &  8 & 9.08 & 11.61 \\
Xie et al \cite{Xie}        & 19 &  7 & 8.26 & 10.12 \\
Wiesenfeld \cite{vocho}     & 35 &  9 & 8.80 & 10.21 \\
Halonen \cite{Halonen}      & 39 & 24 & 0.99 &  1.58 \\
present                     & 44 & 23 & 0.81 &  1.16 \\
& & & & \\
\hline
\end{tabular}

\end{table}

\end{document}